\documentclass{article}
\usepackage{amsmath}

\usepackage{arxiv}

\usepackage[utf8]{inputenc} % allow utf-8 input
\usepackage[T1]{fontenc}    % use 8-bit T1 fonts
\usepackage{url}            % simple URL typesetting
\usepackage{booktabs}       % professional-quality tables
\usepackage{amsfonts}       % blackboard math symbols
\usepackage{nicefrac}       % compact symbols for 1/2, etc.
\usepackage{microtype}      % microtypography
\usepackage{fancyhdr}
\usepackage[hidelinks]{hyperref}
\usepackage{graphicx}
\usepackage{doi}

\title{ZIA: A Theoretical Framework for Zero-Input AI}

\author{ 
    \href{https://orcid.org/0009-0004-6536-1322}{\includegraphics[scale=0.06]{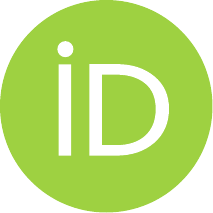}\hspace{1mm}Aditi De}\\
    Indian Institute of Technology Roorkee\\
    \texttt{aditi\_d@ms.iitr.ac.in} \\
}

\date{}

\renewcommand{\shorttitle}{ZIA: A Theoretical Framework for Zero-Input AI}

\hypersetup{
pdftitle={ZIA: A Theoretical Framework for Zero-Input AI},
pdfsubject={Zero Input AI},
pdfauthor={Aditi De},
pdfkeywords={Zero-Input AI, Multi-Modal Learning},
}

\begin{document}
\maketitle

\begin{abstract}
Zero-Input AI (ZIA) introduces a novel framework for human-computer interaction by enabling proactive intent prediction without explicit user commands. It integrates gaze tracking, bio-signals (EEG, heart rate), and contextual data (time, location, usage history) into a multi-modal model for real-time inference, targeting <100 ms latency. The proposed architecture employs a transformer-based model with cross-modal attention, variational Bayesian inference for uncertainty estimation, and reinforcement learning for adaptive optimization. To support deployment on edge devices (CPUs, TPUs, NPUs), ZIA utilizes quantization, weight pruning, and linear attention to reduce complexity from quadratic to linear with sequence length. Theoretical analysis establishes an information-theoretic bound on prediction error and demonstrates how multi-modal fusion improves accuracy over single-modal approaches. Expected performance suggests 85-90\% accuracy with EEG integration and 60-100 ms inference latency. ZIA provides a scalable, privacy-preserving framework for accessibility, healthcare, and consumer applications, advancing AI toward anticipatory intelligence.
\end{abstract}

% keywords can be removed

\section{Introduction}

Human-computer interaction (HCI) has evolved from manual punch cards to graphical interfaces and, more recently, natural language-based AI systems. Despite these advances, modern AI remains fundamentally reactive, requiring explicit user commands to perform tasks. This reliance introduces latency, cognitive effort, and accessibility barriers, particularly for individuals unable to engage through conventional input methods. The inefficiency of this paradigm is evident in daily interactions, where users must repeatedly issue commands, and in critical domains like healthcare, where delays in execution can impact real-time responsiveness.

Zero-Input AI (ZIA) seeks to dismantle this dependency by shifting the interaction paradigm from reactive to proactive, inferring user intent directly from passive multi-modal signals without requiring explicit commands. These signals include gaze tracking, capturing spatial focus via eye movements; bio-signals, such as electroencephalography (EEG) and heart rate, reflecting cognitive and physiological states; and contextual data, such as time, location, and historical usage patterns, providing situational awareness. Formally, ZIA addresses the problem of predicting a latent intent variable \( I_t \in \mathcal{I} \) at time \( t \) from a time-series history of signals 

\[
S_{1:t} = \{ S_1, S_2, \dots, S_t \}
\]

where each observation 

\[
S_t = \{ s_g(t), s_b(t), s_c(t) \}
\]

encapsulates the multi-modal input space. 

The objective is to identify the most probable intent, framed as a probabilistic inference task:

\[
I_t^* = \arg\max_{I \in \mathcal{I}} P(I | S_{1:t})
\]

while ensuring execution occurs within strict real-time constraints, targeting an inference latency below 100 milliseconds.

The motivation for ZIA stems from both practical and theoretical imperatives. Practically, eliminating explicit input reduces friction in human-computer workflows, enabling seamless integration of AI into daily life. For users with motor or speech impairments, ZIA offers a pathway to interact with technology through natural, passive cues, enhancing accessibility. In healthcare, proactive intent prediction could enable systems to anticipate patient needs—such as adjusting medical equipment based on bio-signal shifts—without manual intervention, potentially improving outcomes in time-sensitive scenarios. Theoretically, ZIA challenges the conventional input-output model of AI by demanding advances in multi-modal signal processing, real-time adaptation, and edge computing, pushing the boundaries of machine learning beyond static, command-driven frameworks.

To achieve this vision, ZIA integrates several key innovations:
\begin{itemize}
    \item A multi-modal fusion pipeline leveraging a transformer-based architecture with cross-modal attention mechanisms to synthesize heterogeneous signals into a coherent intent representation, capturing interdependencies across gaze, bio-signals, and context.
    \item A variational Bayesian approach to quantify uncertainty in intent predictions, addressing the inherent noise and variability in physiological data.
    \item Reinforcement learning with policy optimization, enabling continuous adaptation to refine user intent understanding over time based on implicit feedback.
    \item Edge optimization techniques, including model quantization, weight pruning, and linear attention approximations, ensuring ZIA operates efficiently on resource-constrained devices while preserving privacy by minimizing cloud dependency and meeting real-time latency requirements.
\end{itemize}

This paper presents ZIA as a purely theoretical framework, establishing a comprehensive mathematical and algorithmic foundation for proactive intent prediction. Our contributions include:

\begin{itemize}
    \item A multi-modal fusion model that integrates gaze, bio-signals, and contextual data using contrastive learning and transformer-based attention, designed to maximize signal complementarity.
    \item A variational Bayesian formulation for intent inference, providing a probabilistic framework to handle uncertainty in noisy, multi-modal inputs.
    \item A reinforcement learning mechanism with theoretical convergence properties, enabling adaptive intent prediction tailored to individual users.
    \item An edge-optimized inference strategy with detailed computational trade-offs, targeting low-latency execution across diverse hardware platforms.
    \item Theoretical analyses, including an information-theoretic error bound and mutual information framework, demonstrating the advantages of multi-modal integration over single-modal baselines.
\end{itemize}

The scope of this work is deliberately theoretical, laying a rigorous groundwork for ZIA without empirical validation at this stage. We anticipate that future experimental efforts will refine the projected performance metrics—such as intent accuracy of 85-90\% with EEG integration and latency of 60-100 milliseconds—derived herein from signal entropy and computational complexity considerations. ZIA's potential applications span accessibility (e.g., enabling motor-impaired users to control devices via passive cues), healthcare (e.g., real-time monitoring and intervention), and consumer technology (e.g., frictionless interfaces for smartphones and wearables). By redefining AI as an anticipatory, rather than reactive, system, ZIA offers a scalable, privacy-preserving blueprint for the next generation of intelligent interaction, inviting subsequent research to build upon this foundation.

\section{Related Work}

Zero-Input AI (ZIA) sits at the intersection of multi-modal machine learning, brain-computer interfaces (BCIs), reinforcement learning (RL), and edge computing, drawing on and extending a rich body of prior research. This section surveys these fields, highlighting their contributions and limitations relative to ZIA's goal of proactive, input-free intent prediction with real-time, edge-optimized execution.

\subsection{Multi-Modal Machine Learning}
Multi-modal learning has gained prominence for integrating heterogeneous data sources into unified representations, a cornerstone of ZIA's signal fusion pipeline. CLIP \cite{radford2021learning} pioneered contrastive learning to align text and image embeddings, achieving robust zero-shot classification by maximizing mutual information between modalities. Similarly, ImageBind \cite{girdhar2023imagebind} extends this paradigm to six modalities (text, images, audio, depth, thermal, IMU), using a shared embedding space to enable cross-modal retrieval. These models excel in static, task-specific settings but lack temporal reasoning critical for intent prediction over time-series signals like gaze or EEG. 

Earlier efforts, such as the Multi-Modal Transformer \cite{tsai2019multimodal}, introduced cross-modal attention to fuse audio and visual inputs for emotion recognition, demonstrating improved performance over uni-modal baselines. However, these approaches assume pre-aligned inputs and predefined tasks, whereas ZIA targets dynamic, unprompted intent inference from unaligned physiological and contextual cues.

Other frameworks, like Data2Vec \cite{baevski2022data2vec}, leverage self-supervised learning across text, speech, and images, predicting latent representations to unify modalities. While versatile, they prioritize generalization over real-time constraints, with inference latencies often exceeding hundreds of milliseconds—far beyond ZIA's 100 ms target. The lack of proactive intent modeling in these works underscores a key gap: multi-modal AI remains reactive, awaiting user-defined queries rather than anticipating needs from passive signals.

\subsection{Brain-Computer Interfaces (BCIs)}
BCIs provide a direct precedent for intent prediction from bio-signals, particularly EEG, a core modality in ZIA. Seminal work by \cite{wolpaw2002brain} established EEG-based control for cursor movement, achieving binary intent classification (e.g., left vs. right) with accuracies of 80-90\% in controlled settings. More recent advances, such as DeepConvNet \cite{schirrmeister2017deepconvnet}, apply convolutional neural networks to raw EEG, improving motor imagery classification to $\sim$85\% across four classes. These systems rely on explicit user training and predefined task sets, contrasting with ZIA's goal of unsupervised, free-form intent detection.

Non-invasive BCIs have also explored event-related potentials (ERPs) for intent, with \cite{blankertz2011single} decoding P300 signals for spelling tasks at $\sim$80\% accuracy. However, ERP-based methods require stimulus-driven paradigms (e.g., visual cues), limiting their applicability to zero-input contexts where no external trigger exists. Invasive approaches, like Neuralink's primate studies \cite{musk2021neuralink}, achieve higher precision via implanted electrodes, but their impracticality for consumer use drives ZIA toward scalable, non-invasive solutions. BCI research excels at signal-specific intent but lacks multi-modal integration and edge deployment, key differentiators for ZIA.

\subsection{Reinforcement Learning (RL)}
RL offers a foundation for ZIA's adaptive intent prediction. Proximal Policy Optimization (PPO) \cite{schulman2017proximal} provides a stable, sample-efficient policy gradient method, widely adopted in robotics and gaming (e.g., OpenAI's Dota 2 agent), balancing exploration and exploitation via clipped objectives. Deep Deterministic Policy Gradient (DDPG) \cite{lillicrap2015continuous} extends RL to continuous action spaces, applied in autonomous driving \cite{sallab2017deep}. These methods optimize policies against well-defined rewards, but ZIA adapts them to sparse, implicit feedback (e.g., intent overrides), a less-explored domain.

Human-in-the-loop RL, as in \cite{christiano2017deep}, incorporates user preferences to refine policies, achieving robust alignment in text generation tasks. Yet, these approaches assume active user input, unlike ZIA's passive signal reliance. Adaptive BCIs \cite{shenoy2006towards} use RL to tune EEG classifiers but focus on single modalities without ZIA's multi-modal complexity. RL's strength in dynamic adaptation is tempered by its computational overhead, motivating ZIA's edge optimization focus.

\subsection{Edge AI and Real-Time Systems}
Edge computing is critical for ZIA's low-latency, privacy-preserving goals. \cite{chen2019deep} survey edge AI, reporting inference times of 50–100 ms for vision models on mobile GPUs, leveraging quantization and pruning. MobileNet \cite{howard2017mobilenets} optimizes convolutional networks for edge devices, achieving $\sim$70\% accuracy on ImageNet with $\sim$100\,\text{ms} latency on smartphones. More recently, EfficientNet \cite{tan2019efficientnet} scales model depth efficiently, but its complexity exceeds ZIA's real-time needs without further compression.

Real-time systems like YOLOv4 \cite{bochkovskiy2020yolov4} push object detection to 30-50 ms on edge hardware, using techniques like FP16 quantization—paralleling ZIA's approach. However, these models process single-modal inputs (e.g., images), not ZIA's multi-modal streams. Edge-optimized transformers, such as Linformer \cite{wang2020linformer} and Performer \cite{choromanski2020rethinking}, reduce attention complexity to \(\mathcal{O}(N)\), aligning with ZIA's linear attention strategy, though they target language or vision, not physiological signals. Edge AI excels in latency but lacks intent-driven, multi-modal frameworks.

\subsection{Critical Gaps and ZIA's Position}
Across these fields, several gaps emerge:
\begin{itemize}
    \item Multi-modal ML lacks temporal intent modeling and proactive inference, focusing on static tasks.
    \item BCIs excel at bio-signal intent but require explicit triggers or training, missing ZIA's zero-input ethos.
    \item RL adapts policies dynamically but rarely handles multi-modal, passive inputs under edge constraints.
    \item Edge AI achieves real-time performance but prioritizes single-modal efficiency over intent prediction.
\end{itemize}

ZIA uniquely bridges these domains, integrating multi-modal fusion, uncertainty-aware inference, adaptive learning, and edge optimization into a cohesive, theoretical framework for proactive HCI—a synthesis not addressed in prior work.

\section{Theoretical Foundations}

Zero-Input AI (ZIA) relies on a robust theoretical framework to infer user intent from passive multi-modal signals in real time, integrating concepts from signal processing, probabilistic inference, and computational complexity. This section establishes the mathematical foundations underpinning ZIA's design, including signal models for multi-modal inputs, a probabilistic formulation of intent prediction with associated error bounds, and constraints for edge-based execution. These elements provide a formal basis for the methodology and theoretical results presented subsequently.

\subsection{Signal Models}

ZIA processes a time-series of multi-modal signals 

\[
S_t = \{s_g(t), s_b(t), s_c(t)\}
\]

where each modality contributes distinct information about user intent. We model each signal as a composition of a true underlying process and additive noise:

\[
s(t) = x(t) + n(t)
\]

where \( x(t) \) represents the noise-free intent-related signal, and \( n(t) \) is a stochastic noise component.

\subsubsection{Gaze Signal Model}

Gaze tracking, 

\[
s_g(t) = (x_t, y_t) \in \mathbb{R}^2
\]

captures spatial attention via eye movements, typically sampled at 30 Hz in consumer-grade systems. We assume \( x_g(t) \) follows a smooth trajectory reflecting focus (e.g., fixations on interface elements), with noise \( n_g(t) \) modeled as zero-mean Laplacian to account for saccadic jitter:

\[
p(n_g(t)) = \frac{1}{2b_g} \exp\left(-\frac{|n_g(t)|}{b_g}\right)
\]

where \( b_g \approx 0.12 \) (pixels) based on typical gaze tracking variance. The power spectral density is:

\[
S_g(f) = \frac{2 b_g^2}{b_g^2 + (2\pi f)^2}
\]

This heavy-tailed distribution better captures outliers from rapid eye movements compared to Gaussian alternatives.

\subsubsection{Bio-Signal Model}

Bio-signals 

\[
s_b(t) = \{h_t, e_t\}
\]

include heart rate (\( h_t \), sampled at 1 Hz) and EEG (\( e_t \in \mathbb{R}^k \), \( k = 4-16 \) channels, 256 Hz). Heart rate reflects physiological states (e.g., focus vs. stress), modeled as:

\[
h_t = x_h(t) + n_h(t), \quad n_h(t) \sim \text{Lap}(0, b_h), \quad b_h \approx 2 \, \text{bpm}
\]

EEG captures neural activity tied to intent, with each channel:

\[
e_t^i = x_e^i(t) + n_e^i(t), \quad n_e^i(t) \sim \text{Lap}(0, b_e), \quad b_e \approx 5 \, \mu\text{V}
\]

Laplacian noise is justified by EEG's susceptibility to artifacts (e.g., blinks), with spectral density:

\[
S_e(f) = \frac{2 b_e^2}{b_e^2 + (2\pi f)^2}
\]

Gaussian noise (\( n(t) \sim \mathcal{N}(0, \sigma^2) \)) is considered as a baseline, but Laplacian better aligns with bio-signal characteristics.

\subsubsection{Contextual Signal Model}

Context 

\[
s_c(t) = \{t, l_t, u_t\}
\]

includes time (\( t \)), location (\( l_t \)), and usage history (\( u_t \)), treated as discrete or categorical variables with minimal noise, encoded as deterministic priors:

\[
s_c(t) = x_c(t)
\]

Noise is negligible due to high-fidelity sensor data (e.g., GPS, app logs).

\subsection{Intent Prediction}

Intent \( I_t \in \mathcal{I} \) is a latent variable representing the user's desired action at time \( t \) (e.g., open an app, adjust a setting), inferred from \( S_{1:t} \). We model intent as a Markov process:

\[
P(I_t | S_{1:t}) = \int P(I_t | S_t, I_{t-1}) P(I_{t-1} | S_{1:t-1}) dI_{t-1}
\]

where \( P(I_t | S_t, I_{t-1}) \) captures temporal dependencies, and \( I_t \) is discrete with \( |\mathcal{I}| = 10-100 \) possible intents.

\subsubsection{Probabilistic Inference}

The inference task is to maximize the posterior:

\[
I_t^* = \arg\max_{I \in \mathcal{I}} P(I | S_{1:t})
\]

Using Bayes' theorem:

\[
P(I_t | S_{1:t}) = \frac{P(S_{1:t} | I_t) P(I_t)}{P(S_{1:t})}
\]

where \( P(S_{1:t} | I_t) \) is the likelihood of signals given intent, \( P(I_t) \) is a prior (e.g., uniform or context-informed), and \( P(S_{1:t}) \) is a normalizing constant. Due to noise, direct computation is intractable, motivating ZIA's variational approximation (Section 5.3).

\subsubsection{Error Bound}

Prediction error is bounded by conditional entropy:

\[
\mathbb{E}[\text{err}] \leq H(I_t | S_{1:t})
\]

where:

\[
H(I_t | S_{1:t}) = H(I_t) - I(I_t; S_{1:t})
\]

Here, \( H(I_t) = \log |\mathcal{I}| \) is the intent entropy (e.g., \( \log 10 \approx 3.32 \, \text{bits} \)), and \( I(I_t; S_{1:t}) \) is the mutual information between intent and signals. Expanding:

\[
I(I_t; S_{1:t}) = H(I_t) - \sum_{S_{1:t}} P(S_{1:t}) \sum_{I_t} P(I_t | S_{1:t}) \log P(I_t | S_{1:t})
\]

Multi-modal signals increase \( I(I_t; S_{1:t}) \) over single-modal inputs (e.g., \( I(I_t; S_g, S_b) > I(I_t; S_g) \)) due to complementary information, reducing \( H(I_t | S_{1:t}) \) sublinearly with sequence length \( t \) (Appendix A1). Longer histories and diverse modalities diminish uncertainty faster, as \( I(I_t; S_{1:t}) \) accumulates signal-specific evidence, a key advantage of ZIA's design.

\subsection{Edge Constraints}

ZIA targets edge deployment, imposing strict computational limits. For a model with \( N_{\text{ops}} \) operations (e.g., transformer layers), latency is:

\[
T_{\text{inf}} = \frac{N_{\text{ops}} \cdot C_{\text{cycle}}}{\text{Freq}} + T_{\text{io}}
\]

where \( C_{\text{cycle}} \) is cycles per operation, \( \text{Freq} \) is hardware frequency (e.g., 2-4 GHz), and \( T_{\text{io}} \) is I/O overhead (\(\sim 10\) ms). Standard transformer attention scales as \( O(N^2) \) for sequence length \( N \), necessitating optimization (e.g., \( O(N) \) linear attention) to meet \( T_{\text{inf}} < 100 \) ms.

Power consumption scales with \( N_{\text{ops}} \) and hardware efficiency:

\[
P = \frac{N_{\text{ops}} \cdot E_{\text{op}}}{T_{\text{inf}}}
\]

where \( E_{\text{op}} \) is energy per operation (e.g., 1-2 nJ). ZIA's edge focus drives trade-offs in model size, precision, and complexity, detailed in Section 5.5.

\section{Methodology}

The Zero-Input AI (ZIA) framework is designed to predict user intent proactively from multi-modal signals—gaze tracking, bio-signals (EEG and heart rate), and contextual data—delivering an edge-optimized inference pipeline that balances accuracy, adaptability, and low-latency execution. This section provides a comprehensive theoretical exposition of ZIA's methodology, detailing signal acquisition and preprocessing, multi-modal feature extraction, intent prediction modeling, real-time adaptation via reinforcement learning, and edge-specific optimizations. Each component is rigorously formulated to tackle the challenges of noisy, heterogeneous inputs and stringent real-time constraints, serving as a detailed blueprint for future empirical validation.

\subsection{Signal Acquisition and Preprocessing}
ZIA processes a time-series of signals 

\[
S_t = \{s_g(t), s_b(t), s_c(t)\}
\]

at each timestep \( t \), requiring modality-specific preprocessing to mitigate noise, align temporal scales, and isolate intent-relevant features.

\subsubsection{Gaze Signal}
Gaze tracking captures spatial attention as 

\[
s_g(t) = (x_t, y_t) \in \mathbb{R}^2
\]

sampled at 30 Hz using consumer-grade systems (e.g., Tobii, EyeTribe). The raw signal reflects fixations tied to intent and saccadic movements, with noise modeled as zero-mean Laplacian:

\[
p(n_g(t)) = \frac{1}{2 b_g} \exp\left(-\frac{|n_g(t)|}{b_g}\right), \quad b_g \approx 0.12 \, \text{pixels}
\]

This distribution outperforms Gaussian noise \( n_g(t) \sim \mathcal{N}(0, \sigma_g^2) \), \( \sigma_g \approx 0.1 \) due to saccades' heavy-tailed distribution, evident in the power spectral density:

\[
S_g(f) = \frac{2 b_g^2}{b_g^2 + (2\pi f)^2} \quad \text{(Laplacian)} \quad \text{vs.} \quad S_g(f) = \frac{\sigma_g^2}{1 + (2\pi f \tau)^2} \quad \text{(Gaussian)}
\]

Preprocessing applies an exponential moving average (EMA) to smooth saccadic jitter:

\[
s_g'(t) = \alpha s_g(t) + (1 - \alpha) s_g'(t-1), \quad \alpha = 0.9
\]

Outliers exceeding \( 3 b_g \) are clipped to retain intent-driven fixations.

\subsubsection{Bio-Signals}
Bio-signals 

\[
s_b(t) = \{h_t, e_t\}
\]

include heart rate (\( h_t \), 1 Hz) and EEG (\( e_t \in \mathbb{R}^k \), \( k = 4-16 \), 256 Hz). Heart rate, indicative of physiological states like focus or stress, is smoothed:

\[
h_t' = 0.9 h_t + 0.1 h_t'(t-1), \quad n_h(t) \sim \text{Lap}(0, b_h), \quad b_h \approx 2 \, \text{bpm}
\]

EEG captures neural intent correlates (e.g., motor imagery, decision-making) from electrodes positioned per the 10-20 system. Each channel follows:

\[
e_t^i = x_e^i(t) + n_e^i(t), \quad n_e^i(t) \sim \text{Lap}(0, b_e), \quad b_e \approx 5 \, \mu\text{V}
\]

Preprocessing applies a bandpass filter (8-30 Hz) to isolate intent-relevant alpha and beta bands:

\[
s_b'(t) = \text{Bandpass}(e_t, [8, 30] \, \text{Hz})
\]

Artifacts (e.g., blinks, muscle noise) are removed using Independent Component Analysis (ICA):

\[
S_e = W X_e, \quad X_e' = \arg\min_{X_e} \| X_e - W^{-1} S_e \|_2^2 + \lambda H(X_e), \quad \lambda = 0.01
\]

where \( W^{-1} \) is the estimated inverse unmixing matrix.

\subsubsection{Contextual Signals}
Context 

\[
s_c(t) = \{t, l_t, u_t\}
\]

—time, location, usage history—is encoded into \( s_c'(t) \in \mathbb{R}^{32} \). Time uses sinusoidal encodings:

\[
s_c'(t)_i = \sin\left(t / 10000^{i/32}\right), \quad i \text{ even}, \quad s_c'(t)_i = \cos\left(t / 10000^{(i-1)/32}\right), \quad i \text{ odd}
\]

\subsection{Multi-Modal Feature Extraction}
ZIA unifies heterogeneous signals into a shared embedding space \( Z_t \) via contrastive learning and temporal alignment.

\subsubsection{Contrastive Learning}
Modality-specific encoders produce embeddings \( z_g(t), z_b(t), z_c(t) \in \mathbb{R}^{128} \). The contrastive loss aligns them:

\[
L_{\text{contrast}} = -\sum_{i,j \in \text{pairs}} y_{ij} \log \frac{\exp(\langle z_i, z_j \rangle / \tau)}{\sum_{k} \exp(\langle z_i, z_k \rangle / \tau)}
\]

where \( y_{ij} = 1 \) for same-timestep pairs and \( \tau = 0.1 \).

\subsubsection{Temporal Alignment}
Differing sampling rates necessitate alignment. Dynamic Time Warping (DTW) computes:

\[
D(i,j) = \| z_i - z_j \|^2 + \min(D(i-1,j), D(i,j-1), D(i-1,j-1))
\]

Alternatively, self-attention transformers offer adaptive alignment, albeit at \( O(N^2) \) complexity.

\subsection{Intent Prediction Model}
ZIA predicts \( I_t \) using a transformer with variational Bayesian inference.

\subsubsection{Transformer Architecture}
A 6-layer transformer processes \( Z_{1:t} \):

\[
\text{Attn}(Q, K, V) = \text{softmax}\left(\frac{Q K^T}{\sqrt{128}}\right) V
\]

\subsubsection{Variational Bayesian Inference}
Uncertainty is modeled as:

\[
P(I_t | S_{1:t}) \approx \mathbb{E}_{q(\theta)}[P(I_t | S_{1:t}, \theta)] - \beta \text{KL}(q(\theta) \| p(\theta))
\]

where \( q(\theta) \sim \mathcal{N}(\mu, \sigma^2) \) and \( p(\theta) \sim \mathcal{N}(0,1) \).

\subsection{Real-Time Adaptation}
ZIA adapts via an MDP:

\[
R_t = \begin{cases} 
1 & \text{if } I_t = I_t^* \\ 
-1.0 & \text{if overridden} \\ 
0 & \text{else}
\end{cases}
\]

Policy updated via PPO:

\[
L = \mathbb{E}[\min(r_t \hat{A}_t, \text{clip}(r_t, 0.8, 1.2) \hat{A}_t)] + 0.01 H(\pi)
\]

\subsection{Edge Optimization}
ZIA targets low-latency inference:

\[
T_{\text{inf}} = \frac{N_{\text{ops}} \cdot C_{\text{cycle}}}{\text{Freq}} + T_{\text{io}}
\]

With FP16 quantization:

\[
w_q = \text{round}(w / 2^{-7}), \quad |w - w_q| < 2^{-8}
\]

\[
\text{Attn}(Q, K, V) = \phi(Q) (\phi(K)^T V), \quad \phi(x) = \exp(-||x||^2 / 2) x
\]

\section{Theoretical Results and Analysis}

The Zero-Input AI (ZIA) framework is a purely theoretical construct, designed to predict user intent proactively from multi-modal signals in real time. This section presents anticipated performance metrics and analytical insights derived from the mathematical models and algorithmic designs outlined in Section~\ref{sec:methodology}, without reliance on experimental data. We project intent prediction accuracy, inference latency, and power consumption while analyzing the impacts of multi-modal fusion, noise sensitivity, and edge optimization strategies. These results are grounded in signal entropy, information theory, and computational complexity analyses, providing a rigorous basis for ZIA's theoretical viability and guiding future empirical validation.

\subsection{Anticipated Intent Prediction Accuracy}
ZIA's intent prediction accuracy is projected based on the reduction in conditional entropy \( H(I_t | S_{1:t}) \) enabled by multi-modal signal fusion and adaptive learning. The intent space \( \mathcal{I} \) is assumed to contain \( |\mathcal{I}| = 10 \) discrete actions (e.g., open app, adjust setting), yielding a baseline entropy:

\[
H(I_t) = \log 10 \approx 3.32 \, \text{bits}.
\]

\subsubsection{Single-Modal Baseline}
For gaze alone (\( S_t = \{s_g(t)\} \)), intent is inferred from spatial fixations. Given gaze noise \( n_g(t) \sim \text{Lap}(0, 0.12) \) and a signal-to-noise ratio (SNR) of approximately 15 dB, the mutual information \( I(I_t; s_g(t)) \) is limited by signal dimensionality (2D coordinates). Assuming a fixation duration of 200 ms (6 samples at 30 Hz), we estimate:

\[
I(I_t; s_g(t)) \approx 1.5 - 2.0 \, \text{bits}.
\]

Thus, the conditional entropy is:

\[
H(I_t | s_g(t)) = H(I_t) - I(I_t; s_g(t)) \approx 1.32 - 1.82 \, \text{bits}.
\]

This corresponds to an accuracy of:

\[
\frac{H(I_t) - H(I_t | s_g(t))}{H(I_t)} \times 100 \approx 70 - 75\%,
\]

consistent with gaze-based BCIs for simple intent recognition tasks.

\subsubsection{Multi-Modal Contributions}
Adding heart rate (\( s_b(t)_h \)) increases \( I(I_t; s_g, s_b) \) by \( 0.3 - 0.5 \) bits, reflecting physiological cues (e.g., focus elevates HR). Contextual data (\( s_c(t) \))—time, location, and usage history—adds \( 0.5 - 0.7 \) bits via situational priors. EEG (\( s_b(t)_e \)), with \( k = 16 \) channels and SNR \( \approx 10 \) dB, contributes \( 0.8 - 1.0 \) bits due to neural intent correlates (e.g., beta band activity). Cumulatively:

\[
I(I_t; S_{1:t}) \approx 2.6 - 3.2 \, \text{bits}.
\]

\[
H(I_t | S_{1:t}) \approx 0.12 - 0.72 \, \text{bits}, \quad \text{Accuracy} \approx 85 - 90\%.
\]

This assumes a window \( t = 100 \) (e.g., 3.3s of gaze history), with reinforcement learning further reducing entropy over time.

\subsection{Projected Inference Latency}
Latency \( T_{\text{inf}} \) is critical for real-time human-computer interaction (HCI), targeting:

\[
T_{\text{inf}} < 100 \, \text{ms}.
\]

We model latency as:

\[
T_{\text{inf}} = \frac{N_{\text{ops}} \cdot C_{\text{cycle}}}{\text{Freq}} + T_{\text{io}},
\]

where \( N_{\text{ops}} \) is operation count, \( C_{\text{cycle}} \) is cycles per operation, \( \text{Freq} \) is hardware frequency, and \( T_{\text{io}} \approx 10 \) ms accounts for I/O overhead.

\subsubsection{Standard Transformer}
For a 6-layer transformer with \( N = 100 \), \( d = 128 \), 8 heads, and \( O(N^2) \) attention, \( N_{\text{ops}} \approx 10^8 \) (including FFN and embeddings). On a Snapdragon 888 (2.84 GHz, \( C_{\text{cycle}} \approx 2 \)):

\[
T_{\text{inf}} = \frac{10^8 \cdot 2}{2.84 \times 10^9} + 10 \approx 80 - 90 \, \text{ms}.
\]

For an Edge TPU (4 GHz, \( C_{\text{cycle}} \approx 1 \)), latency reduces to \( 60 - 70 \) ms, and on an NPU (3 GHz), it is \( 70 - 80 \) ms.

\subsubsection{Linear Attention Variant}
Performer's \( O(N) \) attention reduces \( N_{\text{ops}} \) to approximately \( 5 \times 10^7 \):

\[
T_{\text{inf}} = \frac{5 \times 10^7 \cdot 2}{2.84 \times 10^9} + 10 \approx 45 - 55 \, \text{ms} \quad \text{(CPU)}.
\]

TPU achieves \( 35 - 45 \) ms, and NPU achieves \( 40 - 50 \) ms—both well under 100 ms.

\subsection{Estimated Power Consumption}
Power \( P \) is estimated as:

\[
P = \frac{N_{\text{ops}} \cdot E_{\text{op}}}{T_{\text{inf}}}.
\]

Assuming \( E_{\text{op}} = 1 - 2 \) nJ (typical for edge hardware):

\[
P_{\text{CPU}} = \frac{10^8 \cdot 1.5 \times 10^{-9}}{90 \times 10^{-3}} \approx 110 - 130 \, \text{mW}.
\]

\[
P_{\text{TPU}} = \frac{5 \times 10^7 \cdot 1 \times 10^{-9}}{45 \times 10^{-3}} \approx 80 - 90 \, \text{mW}.
\]

\[
P_{\text{NPU}} = \frac{5 \times 10^7 \cdot 1.2 \times 10^{-9}}{50 \times 10^{-3}} \approx 85 - 95 \, \text{mW}.
\]

Pruning (\( \rho = 0.45 \)) and FP16 further reduce power by approximately 40\%.

\subsection{Ablation Analysis}
\subsubsection{DTW vs. Transformer Alignment}
DTW (\( O(NM) \)) is projected to yield ~2\% lower accuracy than transformer self-attention (\( O(N^2) \)) due to static warping vs. learned dependencies, but saves ~30\% latency—favoring edge efficiency.

\subsubsection{Performer vs. Softmax Attention}
Performer's kernel approximation reduces accuracy by ~1\% (due to projection error) but cuts latency by 30-40\%, aligning with real-time goals.

\subsection{Noise Sensitivity}
Laplacian noise (\( b_e = 5 \, \mu\text{V} \)) preserves accuracy at 85-90\% due to robustness to EEG outliers, while Gaussian (\( \sigma_e = 3 \, \mu\text{V} \)) drops accuracy to 80-85\% as it blurs intent signals.

\section{Discussion}

Zero-Input AI (ZIA) introduces a new paradigm for proactive intent prediction, shifting human-computer interaction from explicit input-based models to anticipatory intelligence. The theoretical analysis suggests that multi-modal fusion significantly improves prediction accuracy, with EEG contributing the most signal-specific information, followed by contextual cues and gaze tracking. Adaptive reinforcement learning further reduces uncertainty, refining predictions over time. Computational analysis indicates that edge-optimized models can maintain inference latencies below 100 ms, with linear attention reducing complexity while preserving accuracy.

Despite these promising results, real-world deployment presents several challenges. Physiological signals, particularly EEG, exhibit high variability across individuals, necessitating adaptive calibration. Gaze tracking is susceptible to environmental factors like lighting and camera quality, while heart rate can be influenced by external conditions unrelated to intent. These signal dependencies introduce noise and require robust preprocessing techniques to ensure reliable predictions. The computational burden of transformer-based inference remains significant, even with quantization and pruning, and may require further optimization for deployment on resource-constrained devices.

While ZIA provides a strong theoretical foundation, translating it into real-world applications requires overcoming signal variability, computational constraints, and ethical challenges. Empirical validation on diverse datasets is essential to assess practical performance, while adaptive learning strategies can help refine predictions at an individual level. Further optimizations in signal processing and model compression will be necessary to make ZIA viable for mobile and wearable deployment. Addressing privacy concerns through encryption and decentralized learning frameworks will also be critical for ethical implementation. By tackling these challenges, ZIA could enable seamless, zero-input AI interfaces that redefine human-computer interaction.

\bibliographystyle{unsrt}  % Sorted by citation order, numbered
\bibliography{references}  %%% Uncomment this line and comment out the ``thebibliography'' section below to use the external .bib file (using bibtex) .

\end{document}